\documentclass[9pt,twocolumn,twoside]{optica}
\setboolean{shortarticle}{true}
\setboolean{minireview}{false}

\usepackage[labelfont=bf,
   justification=justified,
   format=plain]{caption}
   
\usepackage{nicefrac}
\usepackage{multirow}
\title{Einstein-Podolsky-Rosen paradox in a hybrid bipartite system}

\author[1]{Micha\l{} D\k{a}browski}
\author[1,*]{Micha\l{} Parniak}
\author[1]{Wojciech Wasilewski}

\affil[1]{Institute of Experimental Physics, Faculty of
Physics, University of Warsaw, Pasteura 5, 02-093 Warsaw, Poland}

\affil[*]{Corresponding author: michal.parniak@fuw.edu.pl}

\dates{Compiled \today}

\ociscodes{(020.0020) Quantum optics; (270.5585) Quantum information and processing; (230.1040)
Atomic and molecular physics.}

\doi{\url{http://dx.doi.org/10.1364/optica.XX.XXXXXX}}

\begin{abstract}
\normalsize
Entanglement of light and matter is an essential resource
for effective quantum engineering. In particular, collective states
of atomic ensembles are robust against decoherence while preserving
the possibility of strong interaction with quantum states of light.
While previous approaches
to continous-variable quantum interfaces relied on quadratures of light,
here we present an approach based on spatial structure of light-atom entanglement.
We create
and characterize a 12-dimensional entangled state exhibiting quantum correlations between a photon and an atomic ensemble in position and momentum bases. This state allows us to demonstrate the original Einstein-Podolsky-Rosen (EPR) paradox with two different entities, with an
unprecedented delay time of 6 $\mu$s between generation of entanglement and detection
of the atomic state.
\end{abstract}

\setboolean{displaycopyright}{true}

\begin{document}

\maketitle
\thispagestyle{fancy}
\ifthenelse{\boolean{shortarticle}}{\abscontent}{}

With the rapid development of spatially-resolving single-photon detectors,
spatially structured multidimensional entangled states start to play
a key role in modern quantum science. In particular, they find extensive
applications in emerging fields such as quantum imaging \cite{Bennink2004,Lemos2014},
holography \cite{Chrapkiewicz2015}, computation \cite{Tasca2011}
or quantum-enhanced metrology \cite{Jachura2016}. Moreover, spatially-multiplexed
schemes hold a promise to increase the capacity of quantum channels, essential for quantum key distribution \cite{Walborn2006}. 
An appealing perspective is to apply these ideas to light-atom interfaces, that are capable of storing and processing continous-variable (CV) entanglement \cite{Jensen2010}, critical for novel
quantum cryptography \cite{Goorden:14}, computation \cite{Humphreys2014a}
and imaging \cite{Marino2009} schemes. 

The foundations for these modern
ideas have been laid down by Einstein, Podolsky and Rosen \cite{Einstein1935}
in their famous \emph{Gedankenexperiment} that was designed to prove
the quantum theory is incomplete. They considered an entangled state
of two particles with perfectly correlated positions and anti-correlated
momenta. Such a state exhibits an apparent paradox, namely the product
of conditional variances of positions and momenta violate the Heisenberg inequality
$\Delta x\Delta p_{x}\ge\hbar/2$, suggesting the failure of local
realism. In contrast to what EPR predicted, an experiment conducted
with spin-entangled states \cite{Freedman1972} demonstrated non-local features of quantum mechanics. Later came the demonstrations
of the EPR paradox with CV entangled systems, namely quadratures of
light in optical parametric oscillators \cite{Ou1992,Takei2006,Jensen2010},
four-wave mixing \cite{Marino2009} and\textbf{ }recently spin-quadratures
of a degenerate quantum gas \cite{Peise2015}. Original EPR proposal
with true positions and momenta was realized with photons obtained
in spontaneous parametric down-conversion \cite{Howell2004,Edgar2012,Moreau2014,Schneeloch2013} and spontaneous four-wave mixing \cite{Lee2016},
initiating the trend to explore spatial structure of entangled states
of light.

Here we generate a hybrid bipartite entangled state of a photon and
an atomic spin-wave excitation and demonstrate the EPR paradox with
position and momenta of two different entities. We verify the entanglement
by measuring the coincidence patterns corresponding to modulus-squared
spatial wavefunctions of the state. By studying temporal evolution
and decoherence we find that the EPR entanglement may be stored for
several microseconds, which greatly outstrips the previous approach
based on quadratures of squeezed light and a slow-light medium \cite{Marino2009}. 
Spatial entanglement gives a promise to enhance the capacity of quantum memories \cite{Dai2012,Lan:09,NicolasA2014,Zhang:2016aa} and repeater protocols derived from the Duan-Lukin-Cirac-Zoller (DLCZ) scheme \cite{Duan2001}.

The hybrid entangled state is prepared via a two-mode squeezing Raman
interaction, depicted in Fig. \ref{fig:experiment}, which creates
pairs of Stokes photons and spin-wave excitations with anti-correlated
momenta. A single spin-wave excitation with momentum $\mathbf{P}=(P_{x},P_{y})$
is defined as the state of $N$ atoms created by acting with the $\hat{b}_{\mathbf{P}}^{\dagger}$
operator on the spin-wave vacuum:
\begin{equation}
\hat{b}_{\mathbf{P}}^{\dagger}|0\rangle=\sum_{j=1}^{N}\exp\left(\frac{{i}}{\hbar}\mathbf{P\cdot}\mathbf{R}_{j}\right)|g_{1}\ldots h_{j}\ldots g_{N}\rangle,\label{eq:hamiltonian}
\end{equation}
where $\mathbf{R}_{j}$ is the position of $j$-th atom and $|g\rangle$,
$|h\rangle$ are the two 
metastable states defined in Fig. \hyperref[fig:experiment]{\ref*{fig:experiment}(d)}.
For low probability amplitude $\epsilon$ of entangled pair generation  ($\epsilon\ll1$) we approximate
the spatial structure of the momentum anti-correlated squeezed state by
$|00\rangle+\int\mathrm{d}\mathbf{p}\:\mathrm{d}\mathbf{P}\:\tilde{{\Psi}}(\mathbf{p},\mathbf{P})\:\hat{{a}}_{\mathbf{p}}^{\dagger}\:\hat{{b}}_{\mathbf{P}}^{\dagger}\:|00\rangle$, where
$\hat{{a}}_{\mathbf{p}}^{\dagger}$ is the creation operator for photon
with momentum $\mathbf{p}=(p_{x},p_{y})$, with 
the unnormalized Gaussian-shaped
wavefunction \cite{Moreau2014}: 
\begin{figure}[H]
\centering
\includegraphics[width=\linewidth]{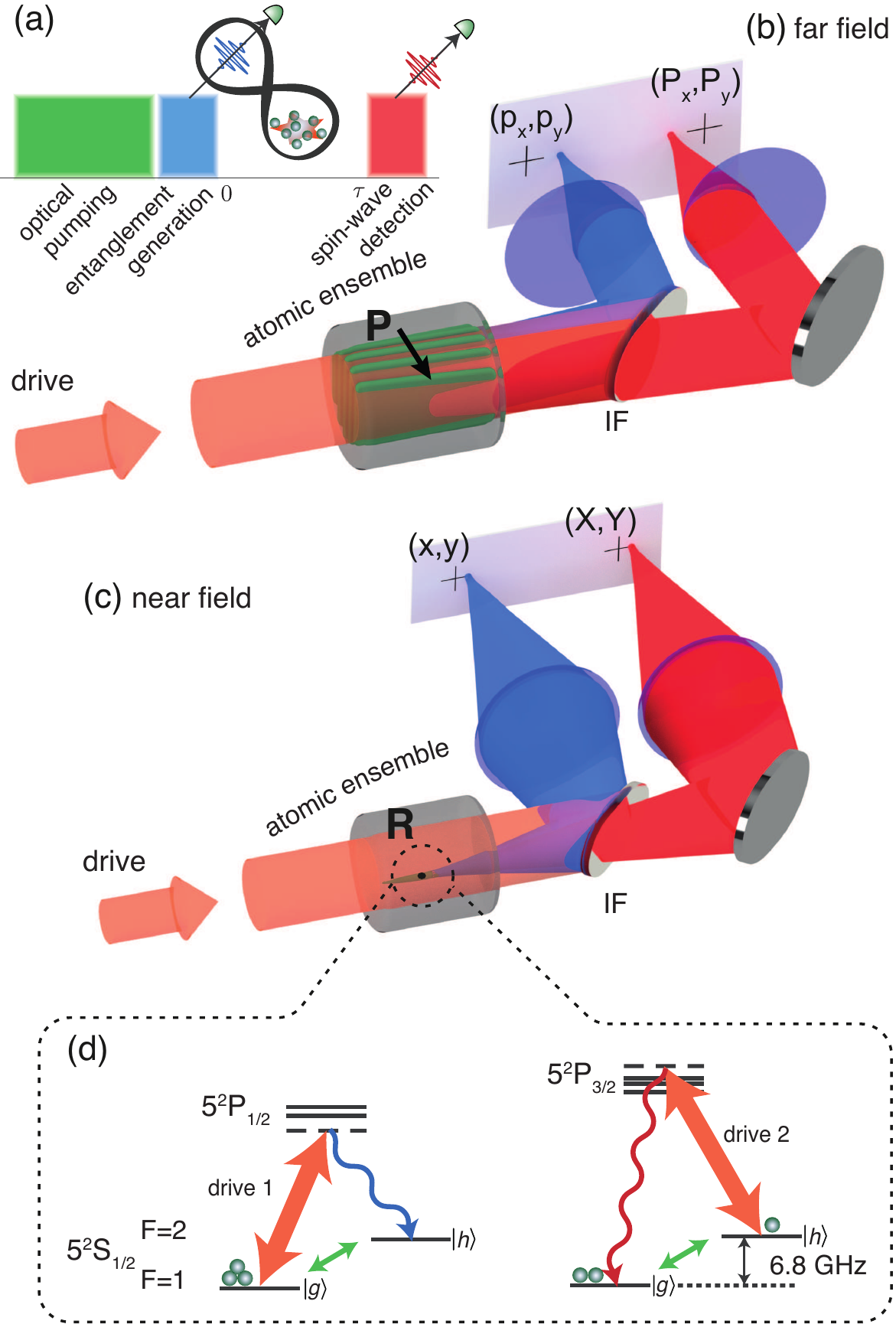}
\caption{ 
(a) The pulse sequence used to generate and then characterize
the hybrid EPR entangled state.
(b) The setup projecting the far field of the atomic ensemble onto
the camera, enabling measurement in the momentum basis, corresponding
to spin-wave-shaped collective atomic excitation. 
(c) An analogous setup imaging the atomic ensemble
onto the detector plane, which enables measurement of positions in the atomic
ensemble from which the photons are emitted.
(d) The relevant atomic level scheme of rubidium-87.  Since the two driving lasers operate at different rubidium
lines, the photons can be separated with an interference filter (IF).
\label{fig:experiment}}
\end{figure}
\noindent
\begin{equation}
\tilde{{\Psi}}(\mathbf{p},\mathbf{P})=\epsilon\frac{\sigma_{-}\sigma_{+}}{\pi}\exp\left(-\sigma_{+}^{2}\frac{|\mathbf{p}+\mathbf{P}|^{2}}{4\hbar^{2}}-\sigma_{-}^{2}\frac{|\mathbf{p}-\mathbf{P}|^{2}}{4\hbar^{2}}\right),\label{eq:momentum}
\end{equation}
in a complete analogy to the biphoton wavefunction.
By Fourier-transforming the wavefunction we obtain a position representation
of the state \cite{Edgar2012,Moreau2014}: 
\begin{equation}
\Psi(\mathbf{r},\mathbf{R})=\epsilon\frac{1}{\pi\sigma_{-}\sigma_{+}}\exp\left(-\frac{|\mathbf{r}-\mathbf{R}|^{2}}{4\sigma_{-}^{2}}-\frac{|\mathbf{r}+\mathbf{R}|^{2}}{4\sigma_{+}^{2}}\right)\label{eq:position}
\end{equation}
 and find that the position of photons $\mathbf{r}=(x,y)$ and spin-wave
excitations $\mathbf{R}=(X,Y)$ are correlated, where $\sigma_{\pm}^{2}=\langle\Delta^{2}|\mathbf{r}\pm\mathbf{R}|\rangle$.

To witness the EPR paradox, we consider the variances of composite
variables, namely sum of momenta
$(p_{x}+P_{x})$ and difference of positions $(x-X)$ from the wavefunctions definitions given by Eqs.
(\ref{eq:momentum}) and (\ref{eq:position}), respectively. As derived by Reid \cite{Reid1989,Reid2009}, the
EPR paradox occurs if the following inequality is satisfied:
\begin{equation}
\langle\Delta^{2}(x-X)\rangle\langle\Delta^{2}(p_{x}+P_{x})\rangle<\hbar^{2}/4.\label{eq:epr}
\end{equation}
However, if the above product is larger than $\hbar^{2}/4$, the
bipartite state may still be entangled. The condition for inseparability is given
by $\langle\Delta^{2}(x-X)\rangle\langle\Delta^{2}(p_{x}+P_{x})\rangle<\hbar^{2}$, as proved by Mancini \textit{et~al.}
\cite{Mancini2002}.
Therefore, we may distinguish three different regimes: the classical
separable regime, the EPR paradox regime, and the intermediate
regime where the correlations, although of quantum origin, may not
be strong enough to demonstrate the EPR paradox but only inseparability
of the state.

\begin{figure}
\includegraphics[width=\linewidth]{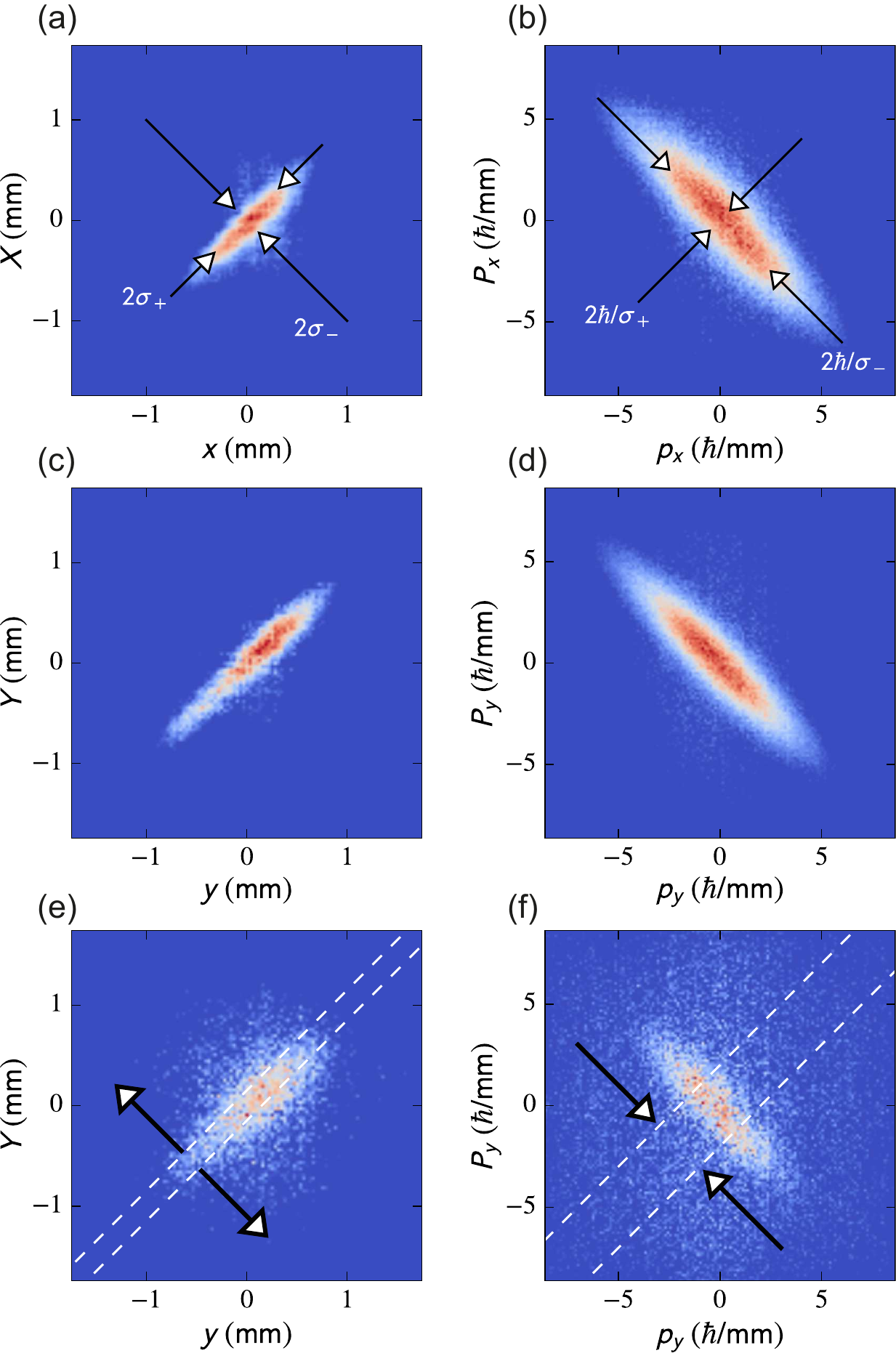}
\centering
\caption{Background-subtracted coincidence maps in positions [(a), (c)] and momenta [(b), (d)] for $x$ [(a), (b)] and $y$ dimensions [(c), (d)] for the shortest delay time $\tau=250$~ns, integrated
over the perpendicular dimension. Correlations in positions and anti-correlations in momenta ale clearly visible.
[(e), (f)] Results precisely corresponding to the ones presented in [(c), (d)], but for a delay time of $\tau=3$ $\mu$s. 
\label{fig:rockets}}
\end{figure}
We now consider the material spin-wave component of the entangled
state. While the photon is sent through a free-space channel at a
long distance, decoherence of the spin-wave excitation occurs. Through
decoherence, a spin wave is lost and the atomic state returns to the
spin-wave vacuum. The rate of this transition is given by $D|\mathbf{P}|^{2}/\hbar^{2}$,
where $D$ is the diffusion coefficient of atoms. As a consequence, the wavefunction from Eq. (\ref{eq:momentum}) is
multiplied by a factor of $\exp(-D|\mathbf{P}|^{2}\tau/\hbar^{2})$
for delay time $\tau$, resulting in a drop of the number of entangled
pairs (see Section S3 of \href{https://www.osapublishing.org/optica/viewmedia.cfm?uri=optica-4-2-272&seq=s001}{Supplement 1}
for derivation). Additionally, the position and momentum variances are changed so that the dimension
of entanglement \cite{Edgar2012}, defined for  delay time $\tau=0$ as $\mathcal{{D}}=(\nicefrac{\sigma_{+}}{\sigma_{-}})^{2}$
drops. Simultaneously the product of variances, essential to witness the EPR paradox, initially rises linearly in time
as $\langle\Delta^{2}|\mathbf{r}-\mathbf{R}|\rangle\langle\Delta^{2}|\mathbf{p}+\mathbf{P}|\rangle\approx\hbar^{2}(\sigma_{-}^{2}+D\tau)/\sigma_{+}^{2}$
. As far as the maximum dimension of entanglement is concerned, we
consider the initial $\sigma_{\pm}^{2}$ variances. Due to decoherence
during the generation and detection operations, which take a finite
total amount of time $T$ to perform, the initial $\sigma_{-}^{2}$
variance is lower-bounded by roughly the mean-squared atomic displacement $DT$. On the other hand, the $\sigma_{+}^{2}$ variance
is upper-bounded by the squared waist radius of the driving beam.

In the experiment we use a 10-cm-long quantum memory vacuum cell containing $N\approx10^{12}$
warm ($75\ ^\circ\mathrm{C}$) rubidium-87 atoms  and krypton at 1 Torr as a buffer gas to make
the atomic motion diffusive. The entanglement is generated by the
driving beam illuminating the ensemble of atoms previously prepared
in the $|g\rangle$ state ($5^{2}S_{1/2},\:F=1$) by optical pumping, which constitutes a spin-wave vacuum. We conduct the experiment
in the spontaneous regime with 0.05 pairs on average generated per single spatial mode in each realization of the experiment, with no amplification due to
build-up of spin-waves. To detect the atomic spin-wave stored inside
atomic ensemble we use stimulated Raman interaction. By sending driving
pulse from another laser after an arbitrary delay time $\tau$, as
seen in Fig. \hyperref[fig:experiment]{\ref*{fig:experiment}(d)}, we perform an on-demand
conversion of a single atomic spin-wave excitation with momentum $\mathbf{P}$ to an anti-Stokes
photon with momentum determined by the phase-matching condition \cite{Chrapkiewicz2016}. All scattered photons are first separated from the driving lasers light and stray fluorescence using a three-stage filtering system and finally registered using state-of-the-art single-photon sensitive
camera with image intensifier (I-sCMOS) \cite{Chrapkiewicz2015}, situated in the near or the far field
which corresponds to the measurement of $(\mathbf{r},\mathbf{R})$
or $(\mathbf{p},\mathbf{P})$ observables, respectively. These two imaging configurations are selectable using flip mirrors (see Section S1 of \href{https://www.osapublishing.org/optica/viewmedia.cfm?uri=optica-4-2-272&seq=s001}{Supplement 1} for more experimental details).

In Fig. \ref{fig:rockets} we present measured bidimensional coincidence
maps corresponding to the modulus-squared wavefunction integrated
over the perpendicular direction. We consider all pairs of registered photons from the two
regions of interest of the camera, mark their positions in the
joint statistics map and at the end subtract the background of accidental
coincidences (see Section S2 of \href{https://www.osapublishing.org/optica/viewmedia.cfm?uri=optica-4-2-272&seq=s001}{Supplement 1} for details on how the data is processed). The coincidence maps clearly demonstrate
correlations in positions and at the same time anti-correlations in
momenta, as expected from the EPR state. After a certain delay time,
the expected effect of diffusional decoherence,
apart from the drop of signal-to-noise ratio, is expansion, or blurring
of the coincidence pattern in the position space. Simultaneously, we observe the shrinkage of
the far-field pattern, corresponding to the decay of high momentum spin-waves, in accordance with the Fourier-transform principle. In turn, the dimension of entanglement $\mathcal{D}$ decreases.

\begin{figure}[H]
\includegraphics[width=\linewidth]{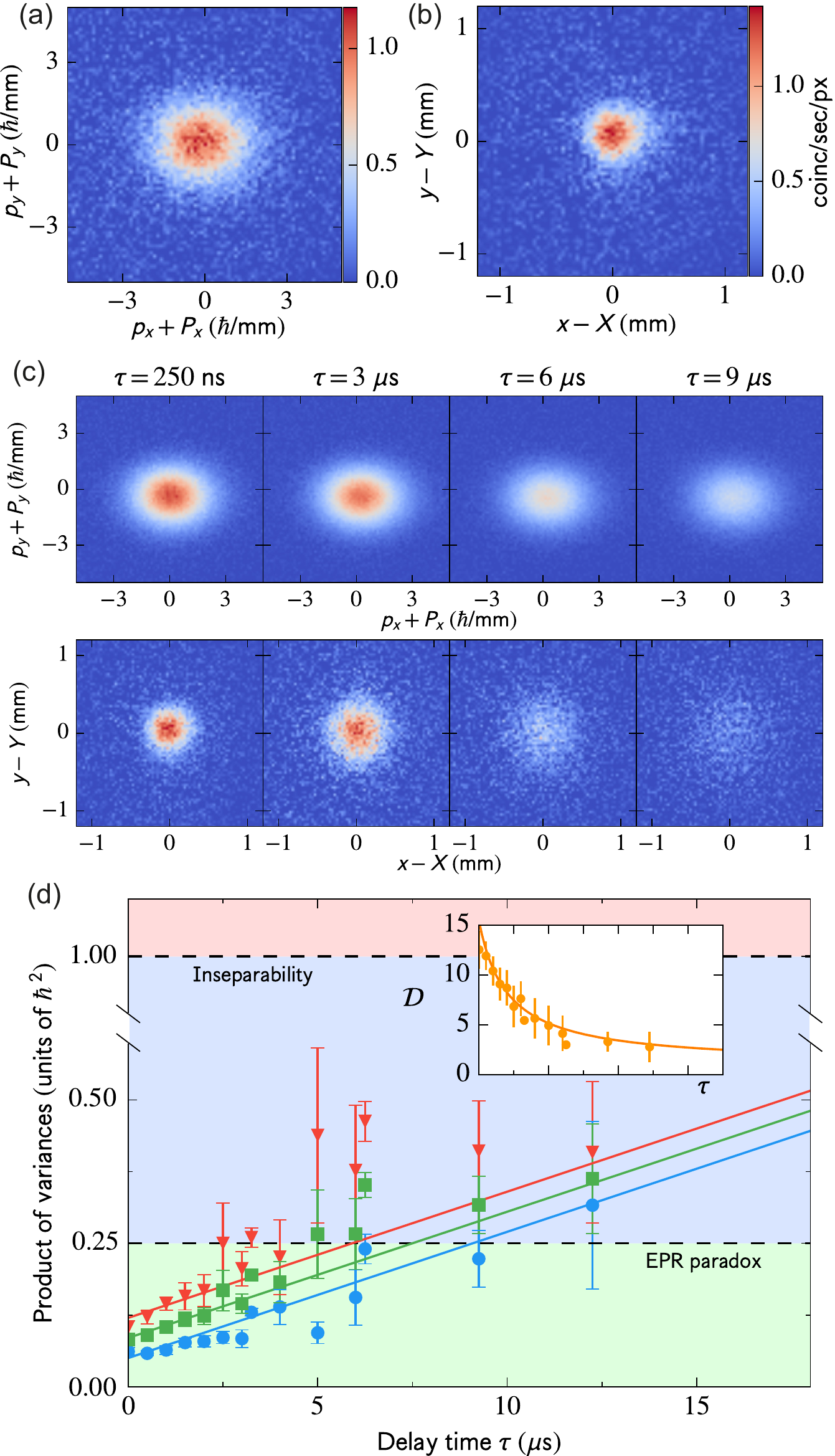}
\centering
\caption{Demonstration of the EPR paradox.
[(a), (b)] Number of 
coincidences in terms of composite variables for the shortest delay
time of $\tau=250$ ns.
(c) The same distributions portrayed
for a set of different delay times. (d) The product of variances. Solid lines correspond to the
expected decay of entanglement due to atomic diffusion for $x$-dimension
(red triangles), $y$-dimension (blue circles) and calculated averages
for the full transverse image \cite{Moreau2012} (green squares).
Inset shows the corresponding dimension of entanglement $\mathcal{{D}}$.
\label{fig:peaks}}
\end{figure}

{To quantify the entanglement of the generated EPR state we 
use the
criterion given by Eq. (\ref{eq:epr}). Similarly as in Fig. \ref{fig:rockets},
we consider the distributions of all registered pairs and subtract
the accidental coincidences. The results presented in Fig.~\ref{fig:peaks} are narrow two-dimensional peaks at zero coordinates for sum of momenta and difference
of positions (see Section S2 of \href{https://www.osapublishing.org/optica/viewmedia.cfm?uri=optica-4-2-272&seq=s001}{Supplement 1} for
cross sections). The width of the peak allows us to estimate the degree
of violation of EPR criterion [Eq. (\ref{eq:epr})] as well as
the amount of entanglement in the system.
We find slightly different $\sigma_{\pm}^{2}$ variances
for the $x$ and $y$-dimensions, which is due to particular arrangement of the experimental setup. Consequently, stronger
degree of entanglement is witnessed in 
 the $y$-dimension where we
observe the EPR paradox until 9 $\mu$s delay time. For the full two-dimensional coincidence distributions, we average
the variances of two-dimensional variables ($\mathbf{r},\mathbf{R}$) and ($\mathbf{p},\mathbf{P}$) for the total position and momentum \cite{Moreau2012}. Finally, for the shortest
delay time of
$\tau=250$~ns we obtain the results presented in \unskip\parfillskip 0pt \par}

\begin{table}[H]
\centering
\caption{Values of variances and their products
for $\tau=250$ ns.
\label{tab:variances}}
\begin{tabular}{|c|c|c|}
\hline 
Variances & Experimental values & Product\tabularnewline
\hline 
$\langle\Delta^{2}(x-X)\rangle$ & 0.040(4) $\mathrm{{mm}}^{2}$ & \multirow{2}{*}{0.10(1)$\hbar^{2}$}\tabularnewline
\cline{1-2} 
$\langle\Delta^{2}(p_{x}+P_{x})\rangle$ & 2.6(1) $\hbar^{2}/\mathrm{{mm}}^{2}$ & \tabularnewline
\hline 
$\langle\Delta^{2}(y-Y)\rangle$ & 0.040(4) $\mathrm{{mm}}^{2}$ & \multirow{2}{*}{0.061(6)$\hbar^{2}$}\tabularnewline
\cline{1-2} 
$\langle\Delta^{2}(p_{y}+P_{y})\rangle$ & 1.50(4) $\hbar^{2}/\mathrm{{mm}}^{2}$ & \tabularnewline
\hline 
$\langle\Delta^{2}|\mathbf{r}-\mathbf{R}|\rangle$ & 0.040(3) $\mathrm{{mm}}^{2}$ & \multirow{2}{*}{0.082(6)$\hbar^{2}$}\tabularnewline
\cline{1-2} 
$\langle\Delta^{2}|\mathbf{p}+\mathbf{P}|\rangle$ & 2.05(6) $\hbar^{2}/\mathrm{{mm}}^{2}$ & \tabularnewline
\hline 
\end{tabular}
\end{table} 

\noindent
Table \ref{tab:variances} and calculate
the dimension of
entanglement $\mathcal{{D}}=12.2\pm0.9$.
Values and uncertainties are
inferred 
from Gaussian fittings to experimental coincidence maps in
Figs. \hyperref[fig:peaks]{\ref*{fig:peaks}(a)} and \hyperref[fig:peaks]{\ref*{fig:peaks}(b)}.

The model of decoherence with predicted linear rise of the product
of variances fits well to our experimental data, as presented in Fig.
\hyperref[fig:peaks]{\ref*{fig:peaks}(d)} (see Section S3 of \href{https://www.osapublishing.org/optica/viewmedia.cfm?uri=optica-4-2-272&seq=s001}{Supplement 1} for derivation).
In particular, after 6 $\mu$s delay time the dimension of entanglement
$\mathcal{D}$ drops so that we are no longer able to certify the
EPR paradox, although the joint state is still inseparable. Notably,
even though the net storage efficiency drops rapidly with time as
high-momentum spin-waves are lost, we observe that our hybrid bipartite
system is still non-classically correlated in positions and momenta. 

In conclusion, we demonstrate generation and characterization of
a hybrid entangled state of light and matter exhibiting EPR correlations
in real space of continuous position-momentum variables, as in the original EPR proposal. As far as
the EPR entanglement is concerned, our approach turns out to be by far
more robust than the only hitherto performed experiment where time-delayed EPR
correlations were demonstrated \cite{Marino2009}. We achieve two
orders of magnitude longer delay time using quantum memory setup and
significantly stronger violation of EPR inequality~[Eq. (\ref{eq:epr})], with product of variances
3 times below the EPR bound for the full two-dimensional coincidence distribution.
Our discussion of contributing experimental factors gives prospects
to further increase the dimension of entanglement.

The spatially-multimode structure of CV entanglement we generate is essential
in terms of quantum information processing and communication, in particular it may provide significant
enhancement for the DLCZ protocol \cite{Duan2001} or improvement
of photon sources \cite{Chrapkiewicz2016}. Temporally-multimode
solutions \cite{Hosseini2009,Rielander2014} available in very similar
systems ask for the connection with spatial multiplexing and polarization degree of freedom to enable demonstrations
of multimode hyperentanglement \cite{Tiranov2015,Zhang:2016aa} or EPR entanglement
with more than two parties \cite{Schaeff:15,Armstrong2015a}. The
possibility to manipulate the stored atomic state opens new avenues
in the vivid topic of EPR-steering \cite{Sun2014}, thus providing
ways to perform novel tests of the quantum theory.

\paragraph{Funding.}
National Science Centre (Poland) Grants No. \mbox{2011/03/D/ST2/01941},
2015/19/N/ST2/01671 and Polish Ministry of Science and Higher
Education \textquotedblleft Diamentowy Grant\textquotedblright{} Project
No. \mbox{DI2013 011943.}

\paragraph{Acknowledgments.}

We acknowledge inspiring discussions with
K. Banaszek  and R. \L apkiewicz, as well as proofreading of the manuscript by
R. Chrapkiewicz, M. Jachura and M. Lipka. 

\bigskip 
\noindent See \href{https://www.osapublishing.org/optica/viewmedia.cfm?uri=optica-4-2-272&seq=s001}{Supplement 1} for supporting content.

\small
\sffamily
\linespread{0.5}

\clearpage



\begin{thebibliography}{10}
\newcommand{\enquote}[1]{``#1''}
\vspace{-1mm}

\bibitem{Bennink2004}
R.~S. Bennink, S.~J. Bentley, R.~W. Boyd, and J.~C. Howell, Phys. Rev. Lett.
  \textbf{92}, 033601 (2004).

\bibitem{Lemos2014}
G.~B. Lemos, V.~Borish, G.~D. Cole, S.~Ramelow, R.~Lapkiewicz, and
  A.~Zeilinger, Nature \textbf{512}, 409 (2014).

\bibitem{Chrapkiewicz2015}
R.~Chrapkiewicz, M.~Jachura, K.~Banaszek, and W.~Wasilewski, Nat. Photonics
  \textbf{10}, 576 (2016).

\bibitem{Tasca2011}
D.~S. Tasca, R.~M. Gomes, F.~Toscano, P.~H. {Souto Ribeiro}, and S.~P. Walborn,
  Phys. Rev. A \textbf{83}, 052325 (2011).

\bibitem{Jachura2016}
M.~Jachura, R.~Chrapkiewicz, R.~Demkowicz-Dobrza{\'n}ski, W.~Wasilewski, and
  K.~Banaszek, Nat. Commun. \textbf{7}, 11411 (2016).

\bibitem{Walborn2006}
S.~P. Walborn, D.~S. Lemelle, M.~P. Almeida, and P.~H.~S. Ribeiro, Phys. Rev.
  Lett. \textbf{96}, 090501 (2006).

\bibitem{Jensen2010}
K.~Jensen, W.~Wasilewski, H.~Krauter, T.~Fernholz, B.~M. Nielsen, M.~Owari,
  M.~B. Plenio, A.~Serafini, M.~M. Wolf, and E.~S. Polzik, Nat. Phys.
  \textbf{7}, 13 (2010).

\bibitem{Goorden:14}
S.~A. Goorden, M.~Horstmann, A.~P. Mosk, B.~\v{S}kori\'{c}, and P.~W.~H.
  Pinkse, Optica \textbf{1}, 421 (2014).

\bibitem{Humphreys2014a}
P.~C. Humphreys, W.~S. Kolthammer, J.~Nunn, M.~Barbieri, A.~Datta, and I.~A.
  Walmsley, Phys. Rev. Lett. \textbf{113}, 130502 (2014).

\bibitem{Marino2009}
A.~M. Marino, R.~C. Pooser, V.~Boyer, and P.~D. Lett, Nature \textbf{457}, 859
  (2009).

\bibitem{Einstein1935}
A.~Einstein, B.~Podolsky, and N.~Rosen, Phys. Rev. \textbf{47}, 777 (1935).

\bibitem{Freedman1972}
S.~J. Freedman and J.~F. Clauser, Phys. Rev. Lett. \textbf{28}, 938 (1972).

\bibitem{Ou1992}
Z.~Y. Ou, S.~F. Pereira, H.~J. Kimble, and K.~C. Peng, Phys. Rev. Lett.
  \textbf{68}, 3663 (1992).

\bibitem{Takei2006}
N.~Takei, N.~Lee, D.~Moriyama, J.~S. Neergaard-Nielsen, and A.~Furusawa, Phys.
  Rev. A \textbf{74}, 060101 (2006).

\bibitem{Peise2015}
J.~Peise, I.~Kruse, K.~Lange, B.~L{\"{u}}cke, L.~Pezz{\`{e}}, J.~Arlt,
  W.~Ertmer, K.~Hammerer, L.~Santos, A.~Smerzi, and C.~Klempt, Nat. Commun.
  \textbf{6}, 8984 (2015).

\bibitem{Howell2004}
J.~C. Howell, R.~S. Bennink, S.~J. Bentley, and R.~W. Boyd, Phys. Rev. Lett.
  \textbf{92}, 210403 (2004).

\bibitem{Edgar2012}
M.~P. Edgar, D.~S. Tasca, F.~Izdebski, R.~E. Warburton, J.~Leach, M.~Agnew,
  G.~S. Buller, R.~W. Boyd, and M.~J. Padgett, Nat. Commun. \textbf{3}, 984
  (2012).

\bibitem{Moreau2014}
P.-A. Moreau, F.~Devaux, and E.~Lantz, Phys. Rev. Lett. \textbf{113}, 160401
  (2014).

\bibitem{Schneeloch2013}
J.~Schneeloch, P.~B. Dixon, G.~A. Howland, C.~J. Broadbent, and J.~C. Howell,
  Phys. Rev. Lett. \textbf{110}, 130407 (2013).

\bibitem{Lee2016}
J.-C. Lee, K.-K. Park, T.-M. Zhao, and Y.-H. Kim, Phys. Rev. Lett.
  \textbf{117}, 250501 (2016).

\bibitem{Dai2012}
H.-N. Dai, H.~Zhang, S.-J. Yang, T.-M. Zhao, J.~Rui, Y.-J. Deng, L.~Li, N.-L.
  Liu, S.~Chen, X.-H. Bao, X.-M. Jin, B.~Zhao, and J.-W. Pan, Phys. Rev. Lett.
  \textbf{108}, 210501 (2012).

\bibitem{Lan:09}
S.-Y. Lan, A.~G. Radnaev, O.~A. Collins, D.~N. Matsukevich, T.~A.~B. Kennedy,
  and A.~Kuzmich, Opt. Express \textbf{17}, 13639 (2009).

\bibitem{NicolasA2014}
A.~Nicolas, L.~Veissier, L.~Giner, E.~Giacobino, D.~Maxein, and J.~Laurat, Nat.
  Photon. \textbf{8}, 234 (2014).

\bibitem{Zhang:2016aa}
W.~Zhang, D.-S. Ding, M.-X. Dong, S.~Shi, K.~Wang, S.-L. Liu, Y.~Li, Z.-Y.
  Zhou, B.-S. Shi, and G.-C. Guo, Nat. Commun. \textbf{7}, 13514 (2016).

\bibitem{Duan2001}
L.-M. Duan, M.~D. Lukin, J.~I. Cirac, and P.~Zoller, Nature \textbf{414}, 413
  (2001).

\bibitem{Reid1989}
M.~D. Reid, Phys. Rev. A \textbf{40}, 913 (1989).

\bibitem{Reid2009}
M.~D. Reid, P.~D. Drummond, W.~P. Bowen, E.~G. Cavalcanti, P.~K. Lam, H.~A.
  Bachor, U.~L. Andersen, and G.~Leuchs, Rev. Mod. Phys. \textbf{81}, 1727 (2009).

\bibitem{Mancini2002}
S.~Mancini, V.~Giovannetti, D.~Vitali, and P.~Tombesi, Phys. Rev. Lett.
  \textbf{88}, 120401 (2002).

\bibitem{Chrapkiewicz2016}
R.~Chrapkiewicz, M.~D{\k a}browski, and W.~Wasilewski, Preprint
  arXiv:1604.06049  (2016).

\bibitem{Moreau2012}
P.-A. Moreau, J.~Mougin-Sisini, F.~Devaux, and E.~Lantz, Phys. Rev. A
  \textbf{86}, 010101 (2012).

\bibitem{Hosseini2009}
M.~Hosseini, B.~M. Sparkes, G.~H{\'{e}}tet, J.~J. Longdell, P.~K. Lam, and
  B.~C. Buchler, Nature \textbf{461}, 241 (2009).

\bibitem{Rielander2014}
D.~Riel{\"{a}}nder, K.~Kutluer, P.~M. Ledingham, M.~G{\"{u}}ndo{\u g}an,
  J.~Fekete, M.~Mazzera, and H.~de~Riedmatten, Phys. Rev. Lett. \textbf{112},
  040504 (2014).

\bibitem{Tiranov2015}
A.~Tiranov, J.~Lavoie, A.~Ferrier, P.~Goldner, V.~B. Verma, S.~W. Nam, R.~P.
  Mirin, A.~E. Lita, F.~Marsili, H.~Herrmann, C.~Silberhorn, N.~Gisin,
  M.~Afzelius, and F.~Bussi{\`{e}}res, Optica \textbf{2}, 279 (2015).

\bibitem{Schaeff:15}
C.~Schaeff, R.~Polster, M.~Huber, S.~Ramelow, and A.~Zeilinger, Optica
  \textbf{2}, 523 (2015).

\bibitem{Armstrong2015a}
S.~Armstrong, M.~Wang, R.~Y. Teh, Q.~Gong, Q.~He, J.~Janousek, H.-A. Bachor,
  M.~D. Reid, and P.~K. Lam, Nat. Phys. \textbf{11}, 167 (2015).

\bibitem{Sun2014}
K.~Sun, J.-S. Xu, X.-J. Ye, Y.-C. Wu, J.-L. Chen, C.-F. Li, and G.-C. Guo,
  Phys. Rev. Lett. \textbf{113}, 140402 (2014).

\end{thebibliography}
\end{document}